\documentclass[reprint,superscriptaddress,nofootinbib,amsmath,amssymb,prd,floatfix]{revtex4-1}
\pdfoutput=1
\usepackage{graphicx}
\usepackage[dvipsnames]{xcolor}
\usepackage{amsmath}
\allowdisplaybreaks
\usepackage{amssymb}

\usepackage{slashed}
\usepackage[utf8]{inputenc}
\usepackage[colorlinks=true,linkcolor=blue,bookmarksopen,bookmarksnumbered]{hyperref}
\usepackage{color}
\usepackage[normalem]{ulem}
\usepackage{soul}
\usepackage{units}
\usepackage{rotating}
\usepackage{hhline,multirow,tabularx}
\usepackage{bm}
\usepackage{hyperref}
\usepackage[export]{adjustbox}
\usepackage{mdframed}
\usepackage{comment}
\usepackage{natbib}

\def\bea#1\eea{\begin{align}#1\end{align}} 

\newcommand{\bef}{\begin{figure}[htb]\centering}
\newcommand{\eef}{\end{figure}}

\newcommand{\shao}[1]{\marginpar{\footnotesize\textbf{SHAO}}}

\def\<{\langle}
\def\>{\rangle}

\def\cos{\hbox{cos}}
\def\sin{\hbox{sin}}

\usepackage{lipsum}
\begin{document}

\title{Photon induced proton and anti-proton pair production with ultraperipheral heavy ion collisions at RHIC}

\author{Cheng Zhang}
\affiliation{School of Physics, Hangzhou Normal University, Hangzhou, Zhejiang 311121, China}

\author{Li-Mao Zhang}
\affiliation{Department of Physics, Center for Field Theory and Particle Physics, Fudan University, Shanghai, 200433, China}

\author{Ding Yu Shao}
\email{dingyu.shao@cern.ch}
\affiliation{Department of Physics, Center for Field Theory and Particle Physics, Fudan University, Shanghai, 200433, China}
\affiliation{Key Laboratory of
Nuclear Physics and Ion-beam Application (MOE), Fudan University, Shanghai, 200433, China}
\affiliation{Shanghai Research Center for Theoretical Nuclear Physics, NSFC and Fudan University, Shanghai 200438, China}


\begin{abstract}

We investigate proton-antiproton ($p\bar{p}$) pair production via photon-photon fusion in the ultra-peripheral collisions at RHIC, employing a joint impact parameter and transverse momentum dependent formalism. We consider proton exchange, $s$-channel resonance and hand-bag mechanisms, predicting differential distributions of $p\bar p$ production. Our theoretical predictions can be tested against future measurements at RHIC, to enhance our understanding of photon-photon interactions in strong electromagnetic fields.

\end{abstract}

\maketitle

\section{Introduction}\label{sec:intro}

In relativistic heavy-ion collisions, the intense electromagnetic fields generated by the fast-moving ions provide an exceptional environment to study photon-photon interactions, producing particle-antiparticle pairs. These interactions are fundamental for exploring Quantum Electrodynamics (QED) in strong fields, revealing the non-linear properties of QED and the dynamics of the vacuum \cite{Bertulani:1987tz, Baur:2001jj, Bertulani:2005ru, Baur:2007zz}.

Ultra-peripheral collisions (UPCs) of heavy ions, where the nuclei interact predominantly through their electromagnetic fields rather than through strong interactions, are particularly suited for investigating these processes. In UPCs, the strong electromagnetic fields can lead to photon-photon fusion, resulting in the creation of real particle pairs from the vacuum—a process known as the Breit-Wheeler process \cite{Breit:1934zz, CERES:1995vll, Baur:1998ay, Baltz:2007kq, Klein:2016yzr, Klein:2020fmr, Hattori:2020htm, Copinger:2020nyx, Steinberg:2021lfm}. Lepton and anti-lepton pair production, including electron, muon, and tau lepton pairs, has been observed in UPCs at both the LHC \cite{ATLAS:2015wnx, ATLAS:2017sfe, CMS:2018uvs, ATLAS:2018pfw, ALICE:2018ael, CMS:2018erd, ATLAS:2020epq, ATLAS:2022ryk, CMS:2022arf, ATLAS:2022ryk, CMS:2022arf, ALICE:2022bii} and RHIC \cite{STAR:2018ldd, STAR:2018xaj, STAR:2019wlg, Zhou:2022gbh}.

In addition to lepton pair production, the study of proton-antiproton ($p\bar{p}$) pair production in photon-photon fusion has also been a significant topic of interest. Measurements by collaborations such as CLEO \cite{CLEO:1993ahu}, VENUS \cite{VENUS:1997can}, OPAL \cite{OPAL:2002nhf}, L3 \cite{L3:2003gyz}, and Belle \cite{Belle:2005fji} have provided crucial data for understanding the cross sections and angular distributions of the $\gamma\gamma \rightarrow p\bar{p}$ process in $e^+e^-$ collisions. Early theoretical predictions using leading-twist nucleon wave functions have been unable to match experimental data \cite{Farrar:1985gv, Farrar:1988vz, Chernyak:1984bm}, necessitating the development of more comprehensive models. These models now incorporate mechanisms such as proton exchange \cite{Ahmadov:2016sdg, Klusek-Gawenda:2017lgt}, $s$-channel resonance exchanges \cite{Odagiri:2004mn}, and the hand-bag mechanism \cite{Berger:2002vc, Diehl:2002yh}.

Recent studies, such as Ref.~\cite{Klusek-Gawenda:2017lgt}, have included all the above theory ingredients in order to achieve a quantitative description of the $e^+e^-$ data. Besides, they also extended calculations from electron-positron colliders to UPCs at the LHC. These efforts aim to validate theoretical models through precise cross-section measurements and differential distributions under various experimental conditions in $e^+e^-$ collisions, providing deeper insights into $p\bar{p}$ pair production in photon-photon fusion processes. However, the equivalent photon approximation (EPA) \cite{vonWeizsacker:1934nji, Williams:1934ad} used in \cite{Klusek-Gawenda:2017lgt} does not incorporate the joint transverse momenta of the incoming photons and impact parameter dependence, and it also omits initial polarization information. As a result, it lacks detailed information about the location of the particle production process \cite{Vidovic:1992ik, Hencken:1994my} and the ability to predict differential distributions, such as the transverse momentum imbalance and azimuthal asymmetry. Recently, theoretical efforts \cite{Zha:2018tlq, Klein:2018fmp, Klein:2020jom, Wang:2021kxm, Wang:2022gkd, Klusek-Gawenda:2020eja, Brandenburg:2021lnj, Wang:2022gkd, Lin:2022flv, Wang:2022ihj, Shao:2022cly, Klusek-Gawenda:2020eja, Shao:2022stc, Niu:2022cug, Harland-Lang:2023ohq, Luo:2023syp, Xie:2023vfi, Shao:2023zge, Shao:2023bga, Zhao:2022dac, Ma:2023dac, Zhou:2024cju} made along this line turn out to give a rather satisfactory description. 

\begin{figure}[h]
    \centering
    \includegraphics[scale=0.25]{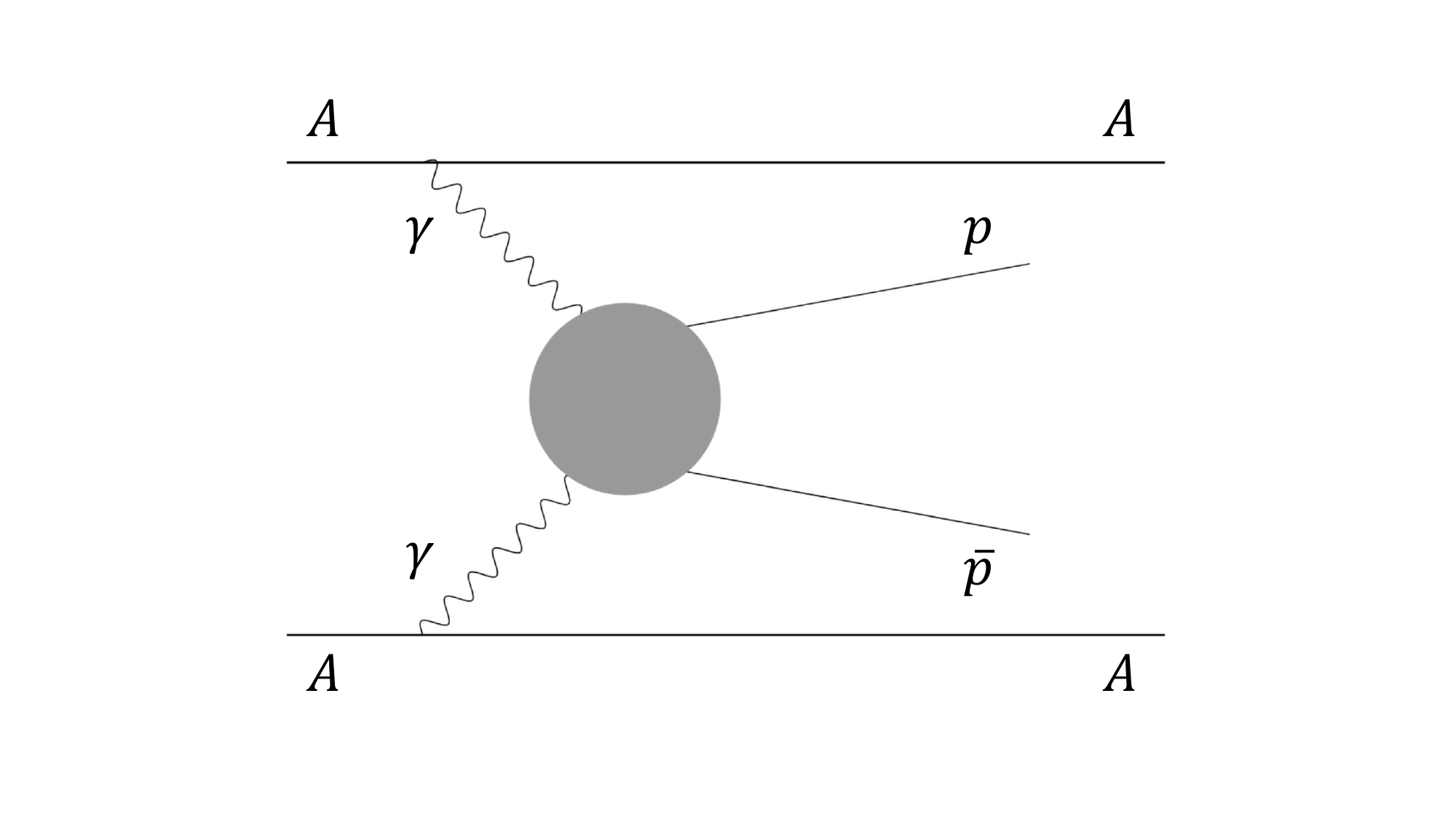}
    \caption{Photon induced proton-antiproton ($p\bar{p}$) pair production from ultraperipheral heavy ion collisions.}
    \label{fig:ppbar}
\end{figure} 

In this paper, we investigate the production of $p\bar{p}$ pairs via photon-photon fusion in UPCs, shown in Fig.~\ref{fig:ppbar}, employing a joint impact parameter and transverse momentum dependent formalism. The impact parameter, defined as the transverse distance between the centers of the colliding nuclei, significantly influences the characteristics of the electromagnetic interactions. This method allows for a precise calculation of the photon-photon fusion cross-section, taking into account the spatial configuration of the colliding ions and the transverse momentum distributions of the virtual photons. By incorporating the initial transverse momenta into our calculations, we can accurately determine differential distributions of the produced $p\bar{p}$ pairs, such as the transverse momentum distribution.

The paper is structured as follows. In the next section, we derive the joint impact parameter and transverse momentum dependent cross section of $p\bar p$ production. We present the numerical result in section \ref{num}. The paper is summarized in section \ref{summary}.

\section{Theoretical formalism}\label{theory}

\begin{widetext}

    \begin{figure}[t]
    \centering
    \includegraphics[scale=0.25]{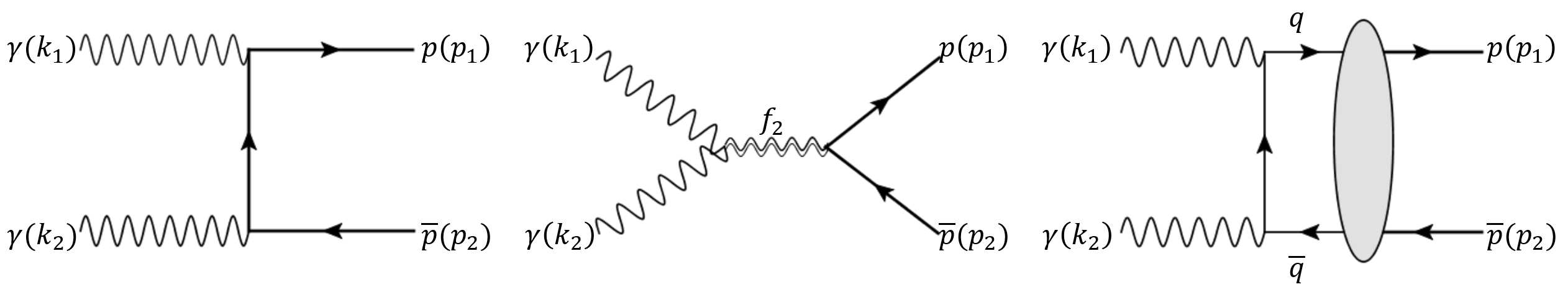}
    \caption{Sample diagrams of proton exchange (left), $s$-channel resonance (middle), and the hand-bag mechanism (right) for the production of $p\bar{p}$ in UPCs.}
    \label{fig:feynman_diagram}
\end{figure} 

\end{widetext}

We employ the EPA method to model the electromagnetic fields of a moving nucleus as a flux of quasi-real photons. This method allows us to approximate the electromagnetic production cross section in UPCs by the cross section for the process $\gamma(k_{1})\gamma(k_{2})\rightarrow p(p_{1})\bar p (p_{2})$, utilizing the equivalent photon distributions. Following the approach outlined in Refs.~\cite{Vidovic:1992ik,Hencken:1994my,Krauss:1997vr}, the cross section can be expressed as
\begin{align}\label{eq:xsec-def}
\sigma=&\int \frac{\mathrm{d}^2 \boldsymbol{b}_{\perp} \mathrm{d}^3 \boldsymbol{p}_1 \mathrm{~d}^3 \boldsymbol{p}_2}{(2 \pi)^3 2 E_1(2 \pi)^3 2 E_2}\\
&\times\Bigg|\int \frac{\mathrm{d}^4 k_1 \mathrm{~d}^4 k_2}{(2 \pi)^4(2 \pi)^4}(2 \pi)^4 \delta^{(4)}\left(k_1+k_2-p_1-p_2\right) \notag\\
&\times \mathcal{M}_{\mu \nu}\left(k_1, k_2, p_1, p_2\right) A_1^\mu\left(k_1, b_{\perp}\right) A_2^\nu\left(k_2, 0\right)\Bigg|^2,\notag
\end{align}
where $\bm{p}_{1,2}$ and $E_{1,2}$ denote the three-momenta and energies of the proton and anti-proton, respectively. Here the bold letters represent vectors in the Euclidean space. In Eq.~\eqref{eq:xsec-def}, $\mathcal{M}_{\mu\nu}$ is the vertex function describing the $\gamma\gamma\rightarrow p\bar p $ process, which is represented by the grey dot in Fig. \ref{fig:ppbar}. The electromagnetic potentials $A_{1}^{\mu}$ and $A_{2}^{\mu}$ are given by
\begin{align}
&A_{1}^{\mu}(k_{1}, b_{\perp})=2\pi Ze\frac{F(-k_{1}^{2})}{-k_{1}^{2}}\delta(k_{1}\cdot u_{1})u_{1}^{\mu}e^{i\bm{k}_{1\perp}\cdot \bm{b}_{\perp}}, \\
&A_{2}^{\mu}(k_{2},0)=2\pi Ze\frac{F(-k_{2}^{2})}{-k_{2}^{2}}\delta(k_{2}\cdot u_{2})u_{2}^{\mu}.
\end{align}
Here the velocities of the heavy ions, represented by $u_{1,2}^{\mu}=\gamma_L(1,0,0,\pm v)$, where $\gamma_L$ is the Lorentz contraction factor. Besides, the nuclear charge density distribution $F(-k^2)$ used in this work is defined by
\begin{align}\label{eq:nucharge}
F\left(\kappa^2\right)=\frac{3\left[\sin \left(\kappa R_A\right)-\kappa R_A \cos \left(R_A\right)\right]}{\left(\kappa R_A\right)^3\left(a^2 \kappa^2+1\right)},
\end{align}
where $a=0.7$ fm and $R_{A}=1.1A^{1/3}$ fm. Substituting the expressions of the electromagnetic potential $A^\mu$ into the cross section \eqref{eq:xsec-def} and simplifying the product of the four-velocities ($u_{1}^\mu$ and $u_2^\nu$) and the vertex function ($\mathcal{M}_{\mu \nu}$), we obtain 
\begin{align}
\sigma \simeq &\frac{(Ze)^{4}}{(4vk^0_1k^0_2)^2}\int\frac{\mathrm{d}^{2}\boldsymbol{b}_{\perp}\mathrm{d}^{3}\boldsymbol{p}_{1}\mathrm{d}^{3}\boldsymbol{p}_{2}}{(2\pi)^{2}E_{1}E_{2}}  \\
&\hspace{-0.5cm}\times \biggl|\int\frac{\mathrm{d}^{2}\boldsymbol{k}_{1\perp}\mathrm{d}^{2}\boldsymbol{k}_{2\perp}}{(2\pi)^{2}(2\pi)^{2}}\delta^{(2)}(\boldsymbol{k}_{1\perp}+\boldsymbol{k}_{2\perp}-\boldsymbol{p}_{1\perp}-\boldsymbol{p}_{2\perp})\notag \\
&\hspace{-0.5cm}\times\mathcal{M}^{ij}(k_{1},k_{2},p_{1},p_{2})e^{i\bm k_{1\perp}\cdot \bm b_{\perp}}k_{1\perp,i}k_{2\perp,j}\frac{F(-k_{1}^{2})}{-k_{1}^{2}}\frac{F(-k_{2}^{2})}{-k_{2}^{2}}\biggr|^{2},\notag 
\end{align}
where $k_{1\perp}$ and $k_{2\perp}$ represent the transverse momentum of the incoming photon, and they are given by 
\begin{align}
    k_{1,2\perp}^\mu=|\boldsymbol{k}_{1,2\perp}|(e^{-i \phi_{\boldsymbol{k}_{1,2\perp}}}\epsilon_{+}^\mu+e^{i \phi_{\boldsymbol{k}_{1,2\perp}}}\epsilon_{-}^\mu)/\sqrt{2},
\end{align}
with the polarization vectors of the photons $\epsilon_{\pm}^\mu\equiv(0,1,\pm i,0)/\sqrt{2}$. In order to obtain the vertex function $\mathcal{M}_{ij}$, in this study, we consider three $p\bar p$ production mechanisms form $\gamma\gamma$ fusion, including the proton exchange, $s$-channel resonance exchanges, and the hand-bag mechanism, which are shown in Fig.~\ref{fig:feynman_diagram}. For convenience, we introduce the Mandelstam variables as
\begin{align} 
& s\equiv\left(k_1+k_2\right)^2=M^2=(P_\perp^2+m_p^2)[2+2\cosh(y_1-y_2)], \notag\\ 
& t\equiv\left(p_1-k_1\right)^2=-P_\perp^2-(P_\perp^2+m_p^2)e^{y_2-y_1}
,\notag\\ 
& u\equiv\left(p_1-k_2\right)^2=-P_\perp^2-(P_\perp^2+m_p^2)e^{y_1-y_2}, 
\end{align}
where $P_\perp=|\bm p_1 - \bm p_2|/2$, $m_p$ represents the proton mass, and $y_1$ and $y_2$ denote the rapidities of proton and anti-proton, respectively. 

For the proton exchange channel, shown in the left panel of Fig.~\ref{fig:feynman_diagram}, the vertex function $\mathcal{M}_{\mu\nu}$ is expressed as

\begin{align}\label{eq:protonQED}
&\mathcal{M}^{\mu\nu}(s_1,s_2)\\
&=  (-i)  \bar{u}\left(s_1,p_1\right)\Bigg[i \Gamma^{\mu}\left(k_1\right) \frac{i\left(p\negthickspace\!\slash_1-k\negthickspace\!\slash_1+m_p\right)}{t-m_p^2+i \epsilon} i \Gamma^{\nu}(k_2) \notag\\
& +i \Gamma^{\nu}\left(k_2\right) \frac{i\left(p\negthickspace\!\slash_1-k\negthickspace\!\slash_2+m_p\right)}{u-m_p^2+i \epsilon} i \Gamma^{\mu}\left(k_1\right)\Bigg] v\left(s_2,p_2\right) F\left(t,u,s\right), \notag
\end{align}
where $s_{1,2} \in \{1/2,-1/2\}$ denote the helicity indices of the proton and anti-proton. Here, $\mathcal{M}_{\mu\nu}$ includes the contribution from both $t$ and $u$ channels, and is expressed in terms of the free proton propagator and the proton electromagnetic vertex 
\begin{align}
    i \Gamma_\mu(k)=-i e\left[\gamma_\mu F_1(k^2)+\frac{i}{2 m_p} \sigma_{\mu \nu} k^\nu F_2(k^2)\right],
\end{align}    
with  $\sigma_{\mu \nu}\equiv\frac{i}{2}\left[\gamma_\mu, \gamma_\nu\right]$. For quasi-real photons in UPCs, we use $F_1(0)=1$ and $F_2(0)=1.793$ for the proton's electromagnetic form factors. To account for the off-shell nature of the proton in the propagator, we introduce a factor $F(t,u,s)$ in Eq.~\eqref{eq:protonQED},  given by \cite{Poppe:1986dq, Klusek-Gawenda:2013rtu, Szczurek:2002bn, Lebiedowicz:2014bea}
\begin{align}
    F(t, u, s)=\frac{\exp \left[2(t-m_p^2)/\Lambda_p^2\right]+\exp \left[2(u-m_p^2)/\Lambda_p^2\right]}{1+\exp\left[-2(s-m_p^2)/\Lambda_p^2\right]},
\end{align}
with $\Lambda_p=1.08$ GeV~\cite{Klusek-Gawenda:2017lgt}, obtained by fitting experimental data in $e^+e^-$ collisions. 

The $s$-channel resonance mechanism involves the fusion of two photons into an $f_2$ meson with $J^{PC}=2^{++}$, which subsequently decays into a proton-antiproton pair, as shown in the middle panel of Fig.~\ref{fig:feynman_diagram}. The helicity amplitudes for this mechanism are given by 
\begin{align} 
\mathcal{M}_{s_1 s_2,\pm\pm}
=&-\frac{1}{2} s^2 \sqrt{s-4 m_p^2} \, \Delta(s)\, a_{f_2 \gamma \gamma} \, g_{f_2 p\bar p}  \\ &\times F^2_{f_2 \gamma \gamma}(s)\left(s-4 m_p^2\right)\left(\cos ^2 \theta-\frac{1}{3}\right) \delta_{s_3 s_4}, \notag \\
\mathcal{M}_{s_1 s_2,\pm\mp} =& -\frac{1}{2} s \sqrt{s-4 m_p^2} \, \Delta(s) b_{f_2 \gamma \gamma}\, g_{f_2 p\bar p}\\
&\times F^2_{f_2 \gamma \gamma}(s)\left(s-4 m_p^2\right) \sin ^2 \theta \delta_{s_3 s_4}, \notag
\end{align}
where we define $\mathcal{M}_{s_1 s_2,\lambda_1 \lambda_2} \equiv \mathcal{M}^{ij}_{s_1 s_2}\epsilon_{\lambda_1i}\epsilon_{\lambda_2j}$, with $\lambda_{1,2}=\pm$ denoting the helicity of the photons. Additionally, we introduce
\begin{align}
    &\Delta(s)\equiv\frac{1}{s-m_{f_2}^2+i m_{f_2} \Gamma_{f_2}}, \\
    &F_{f_2 \gamma \gamma}(s)\equiv\frac{\Lambda_{f_2}^4}{\Lambda_{f_2}^4+(s-m_{f_2}^2)^2}.
\end{align}
Here, $\theta$ is the polar angle of $\boldsymbol{p}_1$ in the center-of-mass frame, with $\cos\theta=\sqrt s/{\scriptstyle \sqrt {s-4m_p^2}} \tanh[(y_1-y_2)/2]$. At the RHIC, the primary contribution is from $f_2(1950)$ resonance, with $m_{f_2}=1944$\,MeV, $\Gamma_{f_2}=473$\,MeV, $\Lambda_{f_2}=1.15$\,GeV, $a_{f_2 \gamma \gamma} g_{f_2 p\bar p}=13.05\,\alpha_e$\,GeV$^{-4}$, $b_{f_2 \gamma \gamma} g_{f_2 p\bar p}=0$ \cite{Klusek-Gawenda:2017lgt}.

The hand-bag mechanism has been investigated in detail in Ref.~\cite{Diehl:2002yh}, showing that the full amplitude can be factorized into a hard factor for $\gamma\gamma\to q\bar q$ and a non-perturbative form factor describing $q\bar q \to p\bar p$ transition, as depicted in the left panel of Fig.~\ref{fig:feynman_diagram}. Explicitly, the helicity amplitudes of the hand-bag mechanics can be expressed as,
\begin{align} 
\mathcal{M}_{s_1 s_2, \lambda_1 \lambda_2}& = \mathcal{A}_{s_1 s_2, \lambda_1 \lambda_2}  \\
 & +\frac{m_p}{\sqrt{s}}\bigg[2 s_1 \mathcal{A}_{-s_1 s_2, \lambda_1 \lambda_2}+2 s_2 \mathcal{A}_{s_1-s_2, \lambda_1 \lambda_2}\bigg], \notag 
\end{align}
with the light-cone amplitude $\mathcal{A}$ given by
\begin{align}
&\mathcal{A}_{s_1 s_2, +-}= -(-1)^{s_1-s_2} \mathcal{A}_{-s_1-s_2,-+}  \\
&=4 \pi \alpha_{e} \frac{s}{\sqrt{t u}}\left[2 s_1 \delta_{s_1,-s_2}\left[r_A(s)+r_P(s)\right]-\frac{\sqrt{s}}{2 m_p} \delta_{s_1 s_2} r_P(s)\right], \notag
\end{align}
where $r_A=0.14$ GeV$^2/s$, $r_P=0.37R_A\sqrt s/2m_p $. In our numerical calculations, we neglect the hand-bag mechanism since it has almost no contribution at RHIC energies, although it should be considered at LHC energies.

Based on the above definition, we obtain the differential cross section written in terms of the helicity amplitudes as
\begin{widetext}
\begin{align}\label{eq:dis}
&\frac{\mathrm{d}\sigma}{\mathrm{d}^{2}\boldsymbol{b}_{\perp}\mathrm{d}^{2}\boldsymbol{P}_\perp\mathrm{d}^{2}\boldsymbol{q}_\perp\mathrm{d}y_1\mathrm{d}y_2} = \frac{Z^{4}\alpha_e^2}{M^{4}}\int\!\frac{[\mathrm{d}\mathcal{K}_{\perp}]}{(2\pi)^8}\!\!\sum_{s_{1},s_{2}} \big[\!\left(\mathcal{M}_{--}\mathcal{M}_{--}^{*}\!+\!\mathcal{M}_{++}\mathcal{M}_{++}^{*}\right)\cos(\phi_{\boldsymbol{k}_{1\perp}}\!\!+\!\phi_{\boldsymbol{k}_{2\perp}}\!\!-\!\phi_{\bar{\boldsymbol{k}}_{1\perp}}\!\!-\!\phi_{\bar{\boldsymbol{k}}_{2\perp}}) \\
&+ \!\left(\mathcal{M}_{-+}\mathcal{M}_{-+}^{*}\!+\!\mathcal{M}_{+-}\mathcal{M}_{+-}^{*}\right)\cos(\phi_{\boldsymbol{k}_{1\perp}}\!\!-\!\phi_{\boldsymbol{k}_{2\perp}}\!\!-\!\phi_{\bar{\boldsymbol{k}}_{1\perp}}\!\!+\!\phi_{\bar{\boldsymbol{k}}_{2\perp}}\!)\!+\! \left(\mathcal{M}_{-+}\mathcal{M}_{+-}^{*}\!+\!\mathcal{M}_{+-}\mathcal{M}_{-+}^{*}\right)\cos(\phi_{\boldsymbol{k}_{1\perp}}\!\!-\!\phi_{\boldsymbol{k}_{2\perp}}\!\!+\!\phi_{\bar{\boldsymbol{k}}_{1\perp}}\!\!-\!\phi_{\bar{\boldsymbol{k}}_{2\perp}}\!)\big],\nonumber
\end{align}
\end{widetext}
where we only keep the azimuthal independent terms, averaging over $\phi_{\boldsymbol{P}_\perp}$ and $\phi_{\boldsymbol{q}_\perp}$. The shorthanded notation is defined as $\mathcal{M}_{\lambda_1 \lambda_2} \equiv \mathcal{M}_{s_1 s_2,\lambda_1 \lambda_2}|_{\phi_{\boldsymbol{p}_{1\perp}}=\phi_{\boldsymbol{p}_{2\perp}}-\pi=0}$ and
\begin{align}
{\cal \int}&[\mathrm{d}{\cal K}_\perp] \equiv \int \mathrm{d}^{2}\bm k_{1\perp}\mathrm{d}^{2}\bm k_{2\perp}\mathrm{d}^{2}\bar {\bm k}_{1\perp}\mathrm{d}^{2}\bar {\bm k}_{2\perp}e^{i(\bm k_{1\perp}-\bar {\bm k}_{1\perp})\cdot \bm b_{\perp}} \notag \\
&\times \delta^{(2)}(\bm k_{1\perp}+\bm k_{2\perp}-\bm q_{\perp}) \delta^{(2)}(\bar {\bm k}_{1\perp}+\bar {\bm k}_{2\perp}-{\bm q}_{\perp}) \notag \\
&\times \mathcal{F}(x_1,{\bm k}_{1\perp}^{2})\mathcal{F}(x_2, {\bm k}_{2\perp}^{2})\mathcal{F}(x_1, \bar {\bm k}_{1\perp}^{2})\mathcal{F}(x_2, \bar {\bm k}_{2\perp}^{2}), 
\end{align}
where $x_1$ and $x_2$ represent the longitudinal momentum fractions of the incoming photons,
\begin{align}\label{eq:energy_frac}
x_{1,2} &= \frac{\sqrt{P_{\perp}^2 + m_p^2}}{\sqrt{s_{N\!N}}}\left(e^{\pm y_{1}} + e^{\pm y_{2}}\right),
\end{align}
with $s_{N\!N}$ being the center-of-mass energy. The transverse momenta of the initial photons in the amplitude and its conjugate amplitude are denoted by $\bm{k}_{1,2\perp}$ and $\bar{\bm{k}}_{1,2\perp}$, respectively. The function ${\cal F}(x_i, \bm{k}_{i\perp}^2)$ describes the amplitude of finding a photon carrying a certain momentum. For a given nuclear charge form factor $F(k^2)$ defined in Eq. \eqref{eq:nucharge}, we have
\begin{align}
    {\cal F}(x, \bm{k}_\perp^2) = |\bm k_{\perp}|\frac{F(\bm{k}_{\perp}^2 + x^2 m_p^2)}{\bm{k}_{\perp}^2 + x^2 m_p^2}.
\end{align}

\section{Phenomenology}\label{num}

In this section, we present our predictions for the production of proton-antiproton pairs at RHIC with $\sqrt{s_{N\!N}}=200$ GeV. The kinematic cuts are chosen as \cite{Wu2023}
\begin{align}
    & P_\perp > 0.2~\text{GeV},\quad |y_{1,2}|<0.5,\notag \\
    & 2.1~\text{GeV} < M < 2.6~\text{GeV}.
\end{align}
Within these kinematic regions, we find that the contribution from the hand-bag mechanism (shown in the right panel of Fig.~\ref{fig:feynman_diagram}) is negligible. Therefore, we exclude its contribution in the following numerical calculations.

\begin{figure}[t]
   \centering\vspace{1cm}
   \includegraphics[scale=0.3]{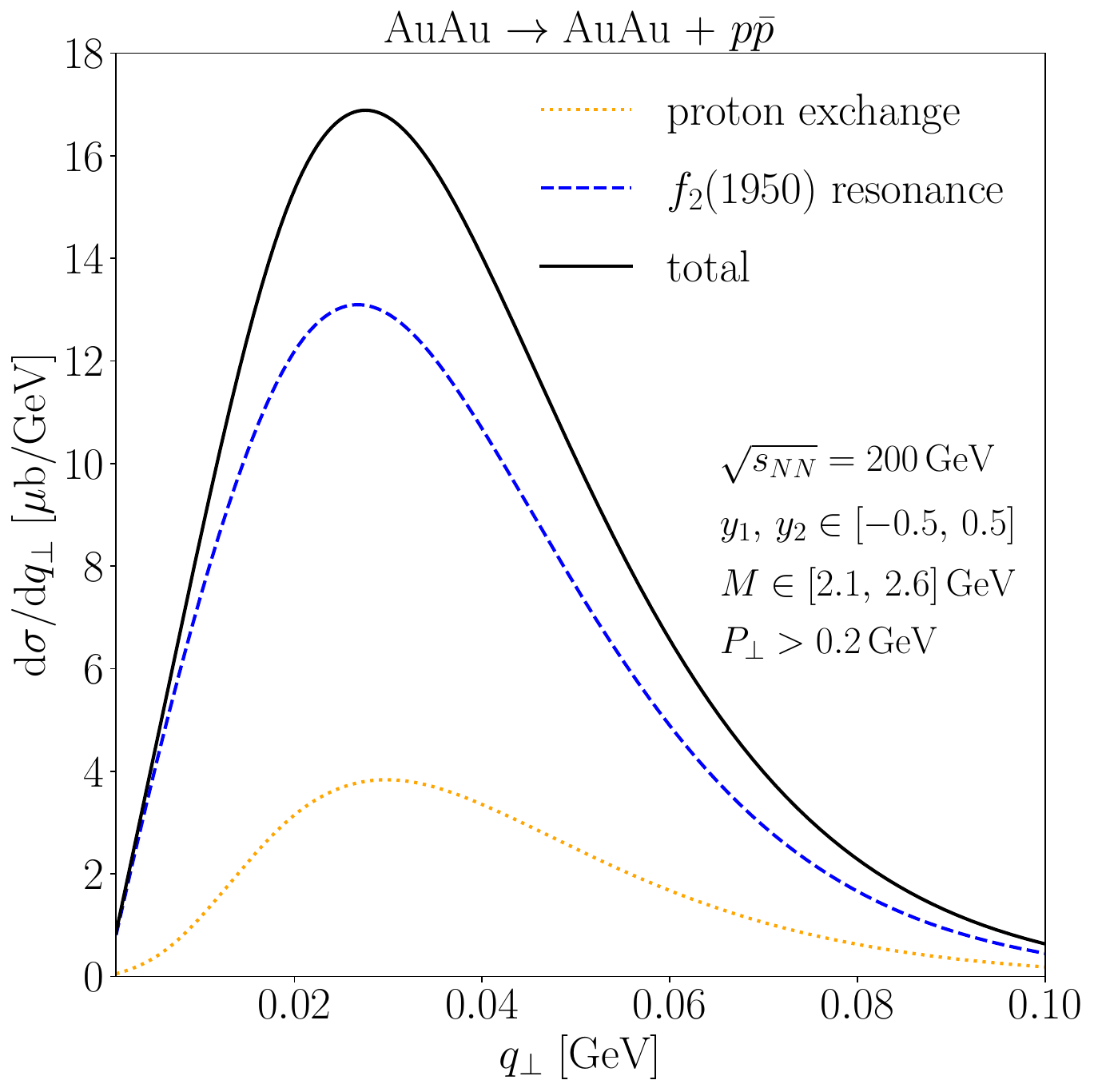} \caption{Total transverse momentum distribution of $p\bar{p}$ pairs at the RHIC with $\sqrt{s_{N\!N}}=200$ GeV. The blue dashed line represents the contribution from the $f_2(1950)$ resonance, while the orange dotted line represents the contribution from proton exchange. The full results are shown as the solid black line. The following kinematic cuts are imposed: the (anti-)proton rapidities $|y_{1,2}|<0.5$, transverse momentum $P_\perp >0.2$ GeV, and the invariant mass of the $p\bar p$ pairs  $2.1~{\rm GeV}<M<2.6~{\rm GeV}$.}
   \label{fig:qt_distribution}
\end{figure}

The total transverse momentum ($q_\perp$) distribution of the produced $p\bar{p}$ pairs is a key observable that provides insight into the photon flux used in our calculations. The photons participating in UPC events are predominantly coherent, with transverse momentum $k_\perp \sim 1/R$ (approximately 30 MeV), where $R$ is the nuclear radius \cite{Li:2019sin, Li:2019yzy}. Fig.~\ref{fig:qt_distribution} shows the $q_\perp$ distribution. The distribution (black line) clearly peaks at low $p_T\sim 30$\,MeV, reflecting the coherent nature of the photons, and exhibits a steep fall-off at higher transverse momenta. Additionally, in Fig.~\ref{fig:qt_distribution}, the blue dashed line represents the contribution from the $f_2(1950)$ resonance, while the orange dotted line represents the contribution from proton exchange. It is obvious that the dominant contributions arise from the $f_2(1950)$ resonance at RHIC. 

\begin{widetext}

\begin{figure}[t]
    \includegraphics[scale=0.25]{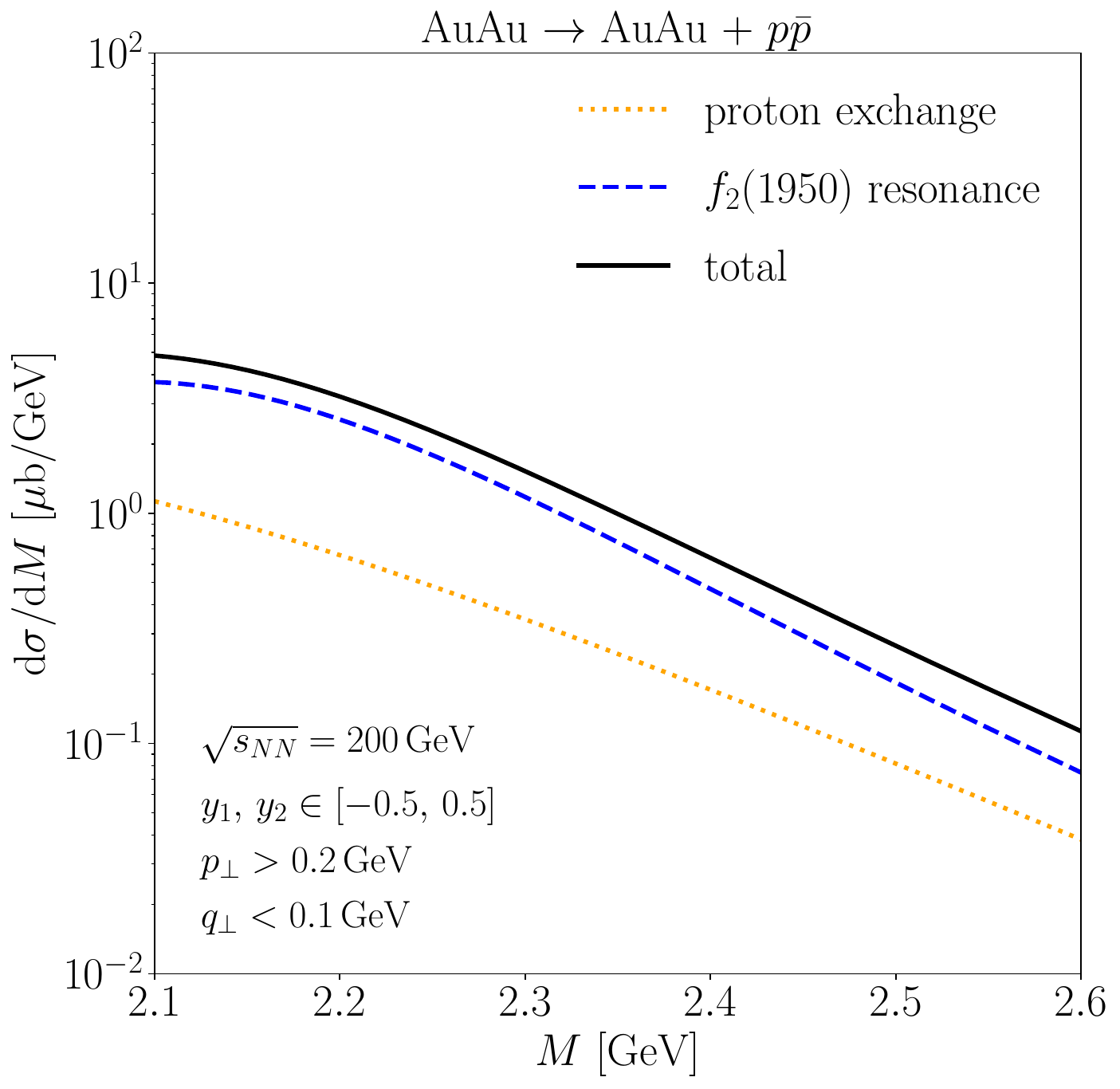} 
    \includegraphics[scale=0.25]{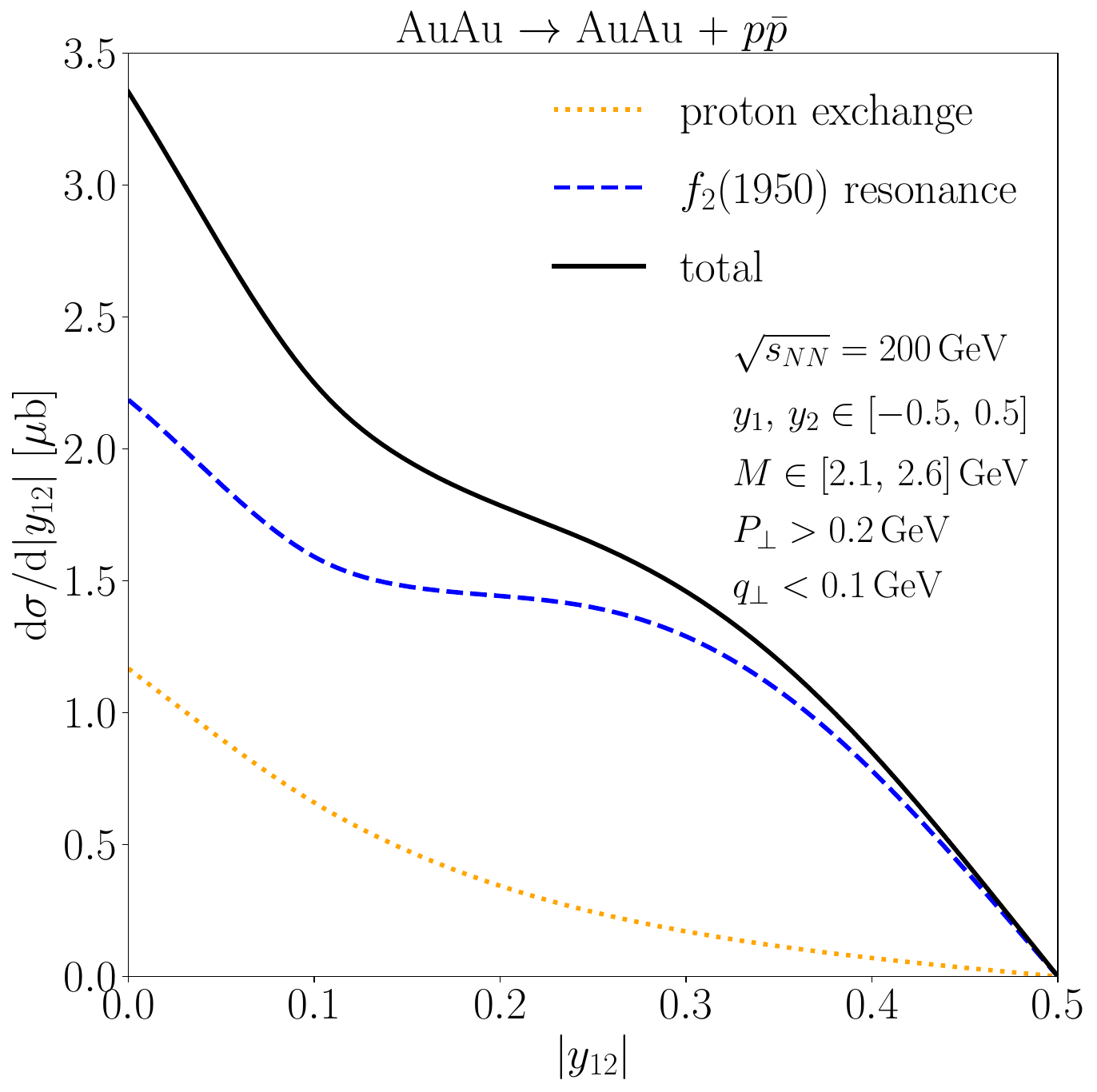} 
    \includegraphics[scale=0.25]{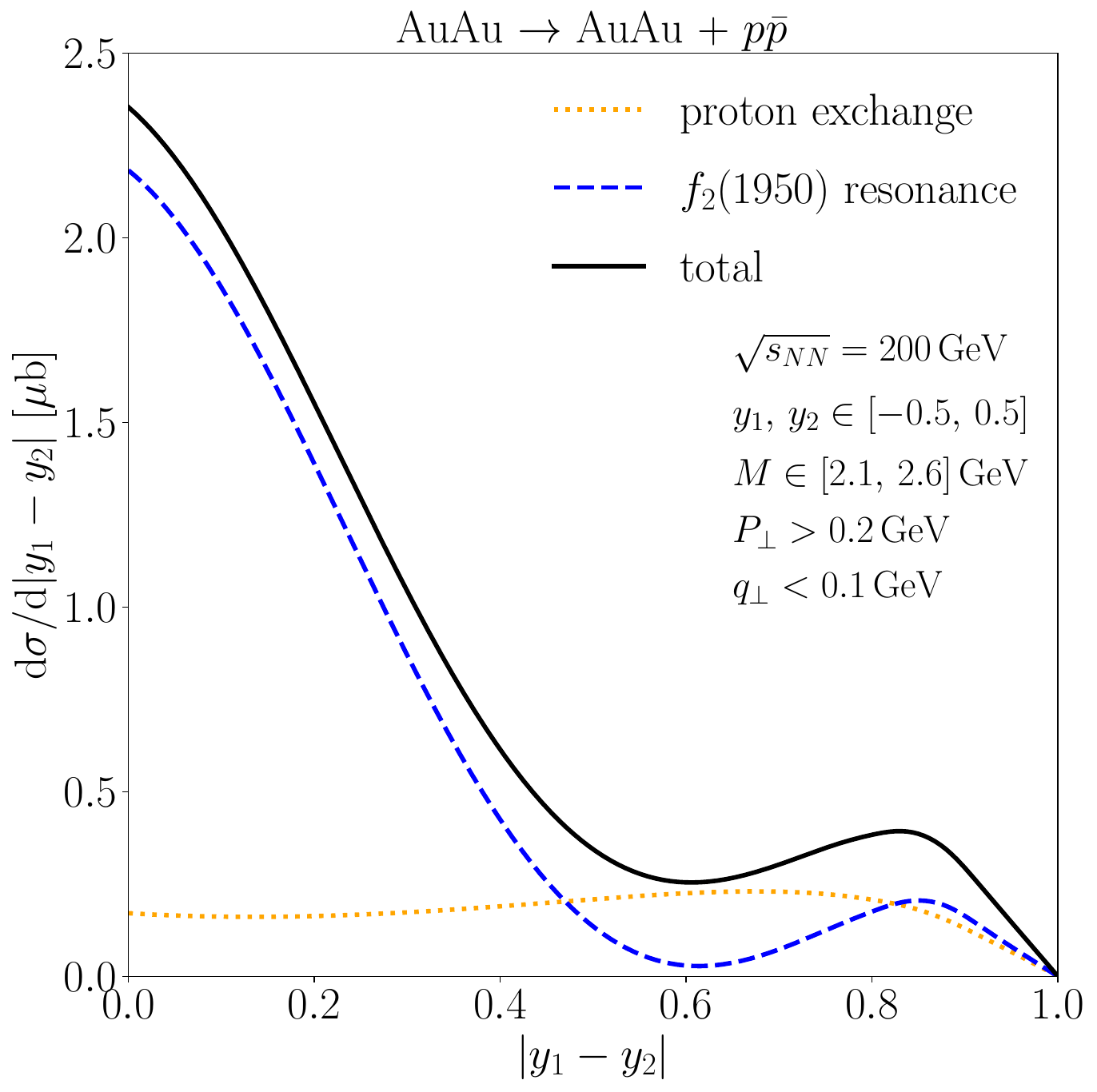}
    \caption{The invariant mass $M$ (left) distribution, total rapidity $|y_{12}|$ (middle) distribution, and rapidity difference $|y_1 - y_{2}|$ (right) distribution for proton-antiproton pairs produced in UPCs at the RHIC. The color schemes follow those in Figure \ref{fig:qt_distribution}, and the kinematic cuts applied are indicated in each panel.}
    \label{fig:dis}
\end{figure}    

\end{widetext}

In addition to the $q_\perp$ distribution, we also study the invariant mass and rapidity distributions using Eq.~\eqref{eq:dis}, where we require the transverse momentum $q_\perp<0.1~{\rm GeV}$. Fig.~\ref{fig:dis} displays the invariant mass $M$ (left), total rapidity $|y_{12}|\equiv|y_1+y_2|/2$ (middle), and rapidity difference $|y_1 - y_2|$ (right) distributions. The blue dashed line represents the contribution from the $f_2(1950)$ resonance, while the orange dotted line represents the contribution from proton exchange. In most regions of invariant mass and rapidity distributions, the dominant contribution comes from the $f_2(1950)$ resonance. 

In the $|y_{12}|$ distribution, the proton exchange curve peaks at lower rapidity values, indicating significant contributions at small rapidities and a sharp decline as $|y_{12}|$ increases. In contrast, the $f_2(1950)$ resonance curve is broader and flatter, showing a more even distribution over a wider total rapidity range. The rapidity difference $|y_1 - y_2|$ distribution also shows that the $f_2(1950)$ resonance significantly contributes to $p\bar{p}$ pairs with similar rapidities, as indicated by the pronounced peak in the blue curve. Conversely, the proton exchange mechanism shows a flatter distribution, contributing more evenly across rapidity differences. This distinction highlights the different dynamics of the two production mechanisms. Future measurements at RHIC will provide a valuable test of different production mechanisms in theoretical predictions.

\section{Conclusions}\label{summary}

In this study, we explored the production of proton-antiproton  pairs via photon-photon fusion in UPCs at the RHIC. By employing a joint impact parameter and transverse momentum dependent formalism, we were able to calculate the cross sections and various differential distributions of $p\bar{p}$ pairs. Our analysis incorporated three mechanisms: proton exchange, $s$-channel resonance exchanges, and the hand-bag mechanism. 

The results from our calculations highlighted the significant contributions of the $f_2(1950)$ resonance and proton exchange mechanisms to the production of $p\bar{p}$ pairs. The transverse momentum distribution of the produced pairs exhibited a clear peak at low $p_T$, indicative of the coherent nature of the photons involved. The $f_2(1950)$ resonance emerged as the dominant contributor in both the invariant mass and rapidity distributions, particularly at small rapidity values. In contrast, the proton exchange mechanism contributed more broadly across a wider range of rapidities, highlighting the different dynamics of the two processes.

Our theoretical framework offers a comprehensive understanding of the $p\bar{p}$ pair production process in UPCs. The findings from this study provide valuable predictions that can be tested against future experimental measurements at RHIC. Such experimental validation will not only refine our current models but also enhance our overall understanding of photon-photon interactions in strong electromagnetic fields.

\vspace{3mm}
\noindent{\it Acknowledgments.}
The authors thank Jin-Hui Chen, Xu-Guang Huang and Jie Zhao for helpful discussion. This work is supported by the National Science Foundations of China under Grant No.~12275052 and No.~12147101.

\bibliography{ref}
\end{document}